\documentclass{elsart}
\usepackage{graphicx}

\begin{document}
\begin{frontmatter}

\title{The CTRW in finance: Direct and inverse problems with some generalizations and extensions}
\author{Jaume Masoliver, Miquel Montero, and Josep Perell\'o}
\address{Departament de F\'{\i}sica Fonamental, 
Universitat de Barcelona, Diagonal, 647, 08028-Barcelona, Spain}
\author{George H. Weiss}
\address{1105 N. Belgrade Rd. Silver Spring, Md. 20902, U.S.A.}

\date{\today}

\begin{abstract}
We study financial distributions within the framework of the continuous time random walk (CTRW). We review earlier approaches and present new results related to overnight effects as well as the generalization of the formalism which embodies a non-Markovian formulation of the CTRW aimed to account for correlated increments of the return.

\end{abstract}
\end{frontmatter}

\section{Introduction}
\label{sec1}

The continuous time random walk (CTRW), first introduced by Montroll and Weiss in 1965 
\cite{montrollweiss}, constitutes a powerful tool for studying the microstructure of many random process appearing in a large variety of physical phenomena. These include transport in disordered media \cite{montroll2,weissllibre}, random networks \cite{berkowitz}, self-organized criticality \cite{boguna}, electron tunneling \cite{gudowska}, earthquake modeling \cite{sornette,grigo1}, hydrology \cite{berkowitz2,dentz} and time-series analysis \cite{grigo2,kutner1}.

Physicists have recently provided some examples of CTRW's applied to finance. Thus the papers by Scalas {\it et al} \cite{scalas} were among the first works addressed to this issue. Further developments have been given in \cite{raberto,scalas-pre} (see \cite{scalas2} for a short review of these works) along with the work of Kutner and Switala \cite{kutner}.

In recent papers \cite{mmw,jebo} we have also developed, in an independent and slightly different way, the formalism of the CTRW designed to study  financial data in the so called ``high frequency regime'', that is, data obtained from tick-by-tick prices. Based on the assumption of independent and identically distributed events, it is possible to get a general expression for the probability distribution of prices at time $t$ in terms of two auxiliary densities that can be estimated from data: the probability density of the pausing time between ticks, $\psi(t)$, and the corresponding density for the magnitude of the price increment at a given tick, $h(x)$ (see Eqs. (\ref{psi}) and (\ref{h}) below for a formal definitions of these quantities). However, with the specific formalism developed in \cite{mmw}, and specially \cite{jebo}, one can only get intraday or at most probabilities at the end of a given session, {\it i.e.}, we have obtained, based on tick-by-tick data, the probability density function for the price {\it within a trading day}. This can be a limitation of the model, since in some  financial applications one is interested in longer distribution of prices. 

In this paper we want to extend some of these developments. In this sense we will present a more general model for the correlation between the densities $\psi(t)$ and $h(x)$ that is based on the theory of copulas \cite{copulas}. We will also present a method to include the so-called ``overnight effect''. Finally, we introduce the general equations for a non-Markovian CTRW which takes into the previous history. 

The paper is organized as follows. In Sect. \ref{sec2} we present a summary of the existent general formalism, and its asymptotic approximations, as applied to financial problems. In Sect. \ref{sec4} we treat the volatility and outline how to treat ``inverse problems'', {\it i.e.,} those of obtaining high-frequency quantities out of the knowledge of low-frequency statistics. In Sect. \ref{sec5} we developed some extensions and generalization of the CTRW which include overnight and memory effects in the description of prices. Conclusions are drawn in Sect. \ref{sec6} and technical details are left to an appendix. 

\section{CTRW: an overview}
\label{sec2}

We present an outline of the CTRW formalism as developed in the framework of financial problems. We refer the reader to Refs. \cite{mmw} and \cite{jebo} for complementary aspects of the subject. 

\subsection{Fundamentals of the Formalism}

The central object of our study is the zero-mean return $X(t)$ defined by
\begin{equation}
X(t)=\ln[S(t+t_0)/S(t_0)]-\langle\ln[S(t+t_0)/S(t_0)]\rangle,
\label{x}
\end{equation}
where $S(t)$ is a speculative price, or the value of an index, and $t_0$ is an initial time. In what follows we will assume stationarity. In this case $X(t)$ is independent of $t_0$. 

We now suppose that $X(t)$ can be described in terms of a CTRW. In this picture $X(t)$ changes at random times $\cdots,t_{-1},t_0,t_1, t_2,\cdots,t_n,\cdots$ and we will first assume that the intervals between successive steps, which we call sojourns or waiting times, $\tau_n=t_n-t_{n-1}$ are independent, identically distributed random variables (in a latter section we will relax this assumption) with a probability density function, $\psi(t)$, defined by 
\begin{equation}
\psi(t)dt={\rm Prob}\{t<\tau_n\leq t+dt\}.
\label{psi}
\end{equation}

At the conclusion of a given sojourn the zero-mean return $X(t)$ undergoes a random change equal to 
$$
\Delta X_n(\tau_n)=X(t_n)-X(t_{n-1})
$$
whose probability density function is defined by
\begin{equation}
h(x)dx={\rm Prob}\{x<\Delta X_n\leq x+dx\}.
\label{h}
\end{equation}
We now define a function $\rho(x,t)$ such that $\rho(x,t)dxdt$ is the joint probability that an increment in return is added whose magnitude, $\Delta X_n$, is between $x$ and $x+dx$ and that the time interval $\tau_n$ between successive jumps is between $t$ and $t+dt$, {\it i.e.,}
\begin{equation}
\rho(x,t)dxdt={\rm Prob}\{x<\Delta X_n<x+dx;\, t<\tau_n<t+dt\}.
\label{rhodef}
\end{equation}
We assume that $\rho(x,t)$ is an even function of $x$ which ensures the absence of drift. Two marginal densities can be formed out of $\rho(x,t)$. The pausing-time density $\psi(t)$ and the density for a random jump $h(x)$. These are related to $\rho(x,t)$ by
\begin{equation}
\psi(t)=\int_{-\infty}^{\infty}\rho(x,t)dx,\qquad 
h(x)=\int_{0}^{\infty}\rho(x,t)dt.
\label{psib}
\end{equation}
Note that a direct consequence of the unbiased assumption, {\it i.e.}: $\rho(-x,t)=\rho(x,t)$, is that $h(x)$ is also symmetric around the origin
\begin{equation}
h(-x)=h(x),
\label{even_h}
\end{equation}
so that its odd moments are equal to zero.

\begin{figure}
\centerline{\includegraphics[width=12cm]{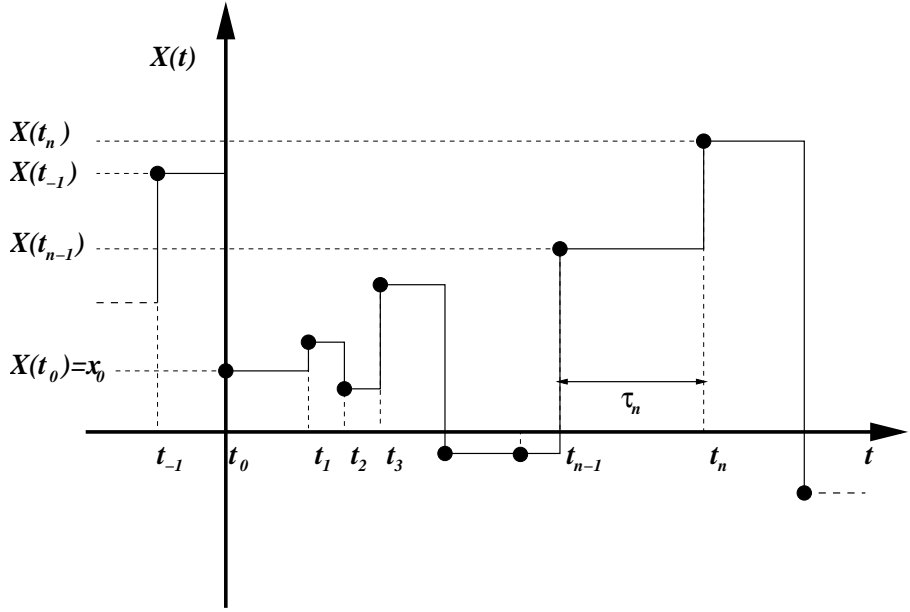}}
\caption{Schematic representation of the return process. The dots mark the value $X(t_n)$ of the return after each sojourn. $\tau_n=t_n-t_{n-1}$ is the time increment of the $n$-th sojourn.}
\label{fig1}
\end{figure}

As to the trajectory of the random process, we assume that $X(t)$ consists of a series of step functions (see Fig. \ref{fig1}). In other words, the return evolves discontinuously and during any sojourn the value of the return remains constant.

The value of the return $X(t)$ at time $t$ will be given by the random value of the height at $t$ (see Fig. \ref{fig1}). Our goal is to obtain the probability density function of $X(t)$. This function is the propagator, defined by
$$
p(x,t)dx={\rm Prob}\{x<X(t)\leq x+dx\}.
$$
From this definition and Fig. \ref{fig1} we see that the propagator prior to the first jump, which we denote by $p_0(x,t)$, is equal to 
\begin{equation}
p_0(x,t)=\Psi(t)\delta(x),
\label{p0}
\end{equation}
where we have assumed that the initial jump occurred at $t=0$, and $\Psi(t)$ is the probability that the time between two ticks is $t$ or greater, 
{\it  i.e.}, 
\begin{equation}
\Psi(t)=\int_t^\infty \psi(t') dt'.
\label{Psi}
\end{equation}

In terms of $p_0$ and $\rho$ we have that the pdf for the return at time $t$, $p(x,t)$, satisfies the following renewal equation:
\begin{equation}
p(x,t)=p_0(x,t)+\int_0^tdt'\int_{-\infty}^{\infty}\rho(x',t')p(x-x',t-t')dx'.
\label{inteq}
\end{equation}
This equation is derived from the consideration that at time $t$, the process is either within the very first sojourn, this given by the first term on the rhs of 
Eq. (\ref{inteq}), or else the first sojourn ended at time $t'<t$, at which time the return 
had value $x'$, and from $(x',t')$ the process is regenerated.

Equation (\ref{inteq}) can be solved in terms of the joint Fourier-Laplace transform 
$$
\tilde{p}(\omega,s)=
\int_0^\infty e^{-st}dt\int_{-\infty}^{\infty} e^{i\omega x}p(x,t)dx.
$$
The solution is
\begin{equation}
\tilde{p}(\omega,s)=\frac{\tilde{p}_0(\omega,s)}{1-\tilde{\rho}(\omega,s)},
\label{hatp}
\end{equation}
where $\tilde{p}_0(\omega,s)$ and $\tilde{\rho}(\omega,s)$ are respectively the joint Fourier-Laplace transforms of the functions $p_0(x,t)$ and $\rho(x,t)$. We can easily see from 
Eqs. (\ref{p0}) and (\ref{Psi}) that the explicit form of $\tilde{p}_0(\omega,s)$ is given by 
\begin{equation}
\tilde{p}_0(\omega,s)=\frac{1-\hat{\psi}(s)}{s},
\label{hatp0}
\end{equation}
where $\hat{\psi}(s)$ is the Laplace transform of the pausing-time density $\psi(t)$. Equations  (\ref{hatp})-(\ref{hatp0}) furnish a complete solution in the transform domain and is a convenient starting point for numerical methods when the inversion cannot be carried out in closed form.

\subsection{Some models}

Unfortunately  the form of $\rho(x,t)$ appearing in the general solution $(\ref{hatp})$ cannot easily be determined from the available data. More easily accessible are the marginal pdf's $h(x)$ and $\psi(t)$. It is therefore essential to assume a functional relation between $\rho(x,t)$ and its marginal densities $\psi(t)$ and $h(x)$. The simplest choice would be based on the assumption that return increments and their duration time are independent random variables \cite{scalas}. In this case
\begin{equation}
\rho(x,t)=h(x)\psi(t),
\label{independent}
\end{equation}
so that Eq. (\ref{hatp}) becomes
\begin{equation}
\tilde{p}(\omega,s)=\frac{[1-\hat{\psi}(s)]/s}
{1-\tilde{h}(\omega)\hat{\psi}(s)}.
\label{independent2}
\end{equation}

However, in some situations one expects some degree of correlation between return increments and their duration, while Eq. (\ref{independent}) implies complete independence between increments and sojourn times. Following that intuition we have presented in \cite{mmw,jebo} two different functional forms for the joint density which take into account the intuitively plausible argument that one must wait for a long time in order for a large variation of return to occur. In the Fourier and Laplace space, these forms are respectively given by 
\begin{equation}
\tilde{\rho}(\omega,s)=\tilde{h}(\omega)\hat{\psi}\left[s\tilde{h}(\omega)\right].
\label{hatrho}
\end{equation}
and
\begin{equation}
\tilde{\rho}(\omega,s)=\hat{\psi}(s)\tilde{h}[\omega\hat{\psi}(s)].
\label{hatrho2}
\end{equation}
That for both models large increments may be infrequent it is clearly seen by the fact that in each case return variations and waiting times are positively correlated. Indeed, let $\tau=t_n-t_{n-1}$ and $\Delta X^2$ be respectively the sojourn time and the return quadratic increment. We define the correlation between these two random variables by the quantity
\begin{equation}
r=\langle\Delta X^2\tau\rangle-\langle\Delta X^2\rangle\langle\tau\rangle.
\label{rcorrelation}
\end{equation}
The cross average $\langle\Delta X^2 \tau\rangle$ can be expressed in terms of $\tilde{\rho}(\omega,s)$ by
\begin{equation}
\langle\Delta X^2 \tau\rangle=\left.
\frac{\partial^3\tilde{\rho}(\omega,s)}{\partial\omega^2\partial s}\right|_{\omega=s=0},
\label{cross_correlation}
\end{equation}
which, after the use of ansatz in Eqs. (\ref{hatrho}) and assuming symmetrical return increments: $\langle\Delta X\rangle=0$, yields 
$\langle\Delta X^2\tau\rangle=2\langle\Delta X^2\rangle\langle\tau\rangle.$ Hence 
$r=\langle\Delta X^2\rangle\langle\tau\rangle>0$. 

For the ansatz in Eq. (\ref{hatrho2}) an analogous calculation gives for the cross correlation 
$\langle\Delta X^2\tau\rangle=3\langle\Delta X^2\rangle\langle\tau\rangle$, whence $r=2\langle\Delta X^2\rangle\langle\tau\rangle>0$. Therefore, for both models $\Delta X^2$ and $\tau$ have a positive correlation, and increasing return variations imply increasing waiting times and vice versa. 

We now concentrate on a third model for the correlation between jumps and waiting times which is borrowed from the theory of copulas: a rather modern development in mathematical statistics which studies probability distributions with fixed marginals \cite{copulas}. In this context, copulas are defined as functions  that join (that is, ``couple") multivariate distribution functions to their one-dimensional distribution functions \footnote{In this sense the models defined by Eqs. (\ref{hatrho}) and (\ref{hatrho2}) can also be termed as copulas.}. Specifically, our model is provided by the so-called ``Clayton copula" in which the joint density $\tilde{\rho}(\omega,s)$ is given by \cite{copulas}
\begin{equation}
\tilde{\rho}(\omega,s)=\left[\tilde{h}^{-\alpha}(\omega)+
\hat{\psi}^{-\alpha}(s)-1\right]^{-1/\alpha},
\label{rhocopula}
\end{equation}
where $\alpha$ is an arbitrary real number. For this model one can easily show that the correlation defined in Eq. (\ref{rcorrelation}) is given by 
$$
r=\alpha \langle\Delta X^2\rangle\langle \tau\rangle.
$$
Therefore any positive correlation between waiting times and jumps will correspond to $\alpha>0$. We should remark at this point that one of the main advantages of the Clayton copula (\ref{rhocopula}) over our previous models is precisely given by the parameter $\alpha$ since it governs the size of the correlation between waiting times and jumps while in (\ref{hatrho}) and (\ref{hatrho2}) this correlation is fixed and solely determined by 
$\langle\Delta X^2\rangle$ and $\langle \tau\rangle$. Moreover, $\alpha$ gives an extra degree of freedom which can be useful in the modeling of empirical data.

The model (\ref{rhocopula}) together with Eq. (\ref{hatp0}) leads us to write the formal solution to the problem given by Eq. (\ref{hatp}) in the following more explicit form:
\begin{equation}
\tilde{p}(\omega,s)=\frac{[1-\hat{\psi}(s)]/s}
{1-\left[\tilde{h}^{-\alpha}(\omega)+\hat{\psi}^{-\alpha}(s)-1\right]^{-1/\alpha}}.
\label{formalsolution}
\end{equation}

\begin{figure}
\centerline{\includegraphics[width=12cm]{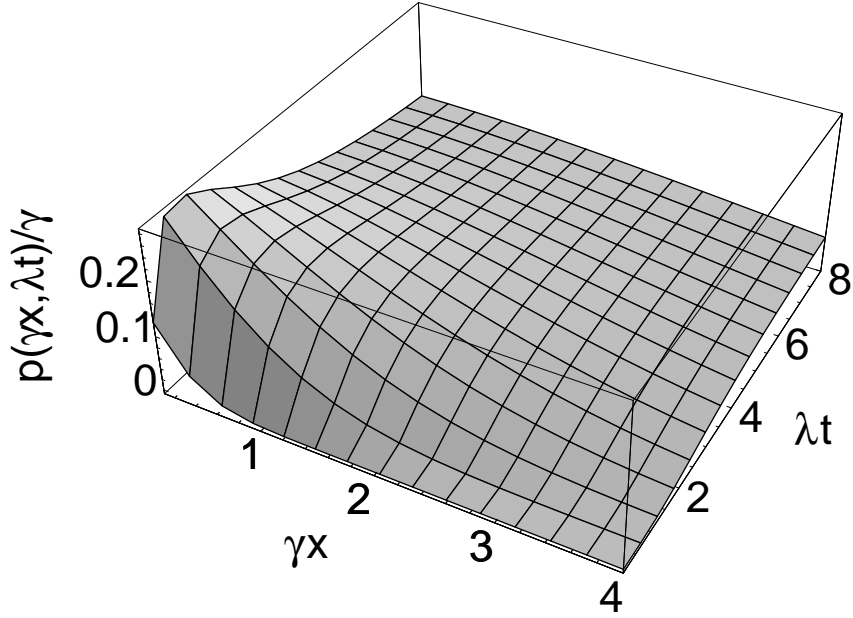}}
\caption{The return pdf in dimensionless units $p(\gamma x,\lambda t)/\gamma$  given in Eq. (\ref{poissonlaplace}) for $x>0$, {\it i.e.}, without the delta function term appearing in (\ref{poissonlaplace}). By the symmetry in $x$, $p(x,t)$ behaves similarly for negative returns.} 
\label{fig1c}
\end{figure}

Let us present an explicit example in which it is possible to evaluate $p(x,t)$ directly. Suppose that $\alpha=1$ and that waiting times and jumps are both exponentially distributed. That is
\begin{equation}
\psi(t)=\lambda e^{-\lambda t},
\label{poissoncase(a)}
\end{equation} 
and 
\begin{equation}
h(x)=\frac{\gamma}{2}e^{-\gamma|x|},
\label{poissoncase}
\end{equation}
where $\gamma>0$ is such that $\langle\Delta X^2\rangle=2/\gamma^{2}$ is the jump variance. In such a case we show in the Appendix \ref{A} that the return pdf is given by the following expression
\begin{equation}
p(x,t)=e^{-\lambda t}\left[\delta(x)+\frac{\gamma}{\sqrt{\pi}}
\int_0^{\sqrt{\lambda t}}e^{\xi^2-\gamma^2x^2/4\xi^2}d\xi\right].
\label{poissonlaplace}
\end{equation}
In Fig. \ref{fig1c} we plot this pdf for positive returns $x>0$. 

We finally remark that models given by Eqs. (\ref{independent}), (\ref{hatrho}),  (\ref{hatrho2}) and (\ref{rhocopula}) are just examples of how to obtain the return pdf, $p(x,t)$, once we have guessed from data plausible forms for $\psi(t)$ and $h(x)$. This is one aspect of the so-called ``direct problem'' and it is certainly the case when we deal with tick-by-tick data and want to obtain an expression for the return pdf at any time. We refer the reader to refs. \cite{mmw}-\cite{jebo} for explicit examples based on actual financial data.

\subsection{Asymptotic analysis}

In the Appendix \ref{B} we show that an asymptotic expression as $t\rightarrow\infty$ of the distribution exemplified by its characteristic function $\tilde{p}(\omega,t)$ is
\begin{equation}
\tilde{p}(\omega,t)\simeq\frac{\langle\tau\rangle}
{\langle\tau e^{i\omega\Delta X}\rangle}
\exp\left\{-\frac{[1-\tilde{h}(\omega)]t}{\langle\tau e^{i\omega\Delta X}\rangle}\right\}
\qquad(t\rightarrow\infty).
\label{asymp}
\end{equation}
We observe that this asymptotic expression is totally general regardless of the correlation between sojourn times $\tau$ and return increments $\Delta X$, requiring only that the mean sojourn time $\langle\tau\rangle$ be finite. Obviously for an usable form of this asymptotic expression we will need to know the correlation between $\tau$ and $\Delta X$. In other words, we will have to specify a form for $\rho(x,t)$ since the correlation appearing in Eq. (\ref{asymp}) is related to $\tilde{\rho}$ by
\begin{equation}
\langle\tau e^{i\omega\Delta X}\rangle=
-\left.\frac{\partial}{\partial s}\tilde{\rho}(\omega,s)\right|_{s=0}.
\label{average}
\end{equation}
Thus, for instance, making use of ansatz in Eq. (\ref{rhocopula}) we have
$$
\langle\tau e^{i\omega\Delta X}\rangle=\langle\tau\rangle\tilde{h}^{1+\alpha}(\omega)
$$
so that
\begin{equation}
\tilde{p}(\omega,t)\simeq\frac{1}{\tilde{h}^{1+\alpha}(\omega)}
\exp\left\{-\frac{[1-\tilde{h}(\omega)]t}
{\langle\tau\rangle\tilde{h}^{1+\alpha}(\omega)}\right\}
\qquad(t\rightarrow\infty).
\label{asympansatz}
\end{equation}

Now assume that $h(x)$ has a finite second moment $\mu_2$ so that  $\tilde{h}(\omega)\simeq 1-\mu_2\omega^2/2$ as $\omega\rightarrow 0$. In this case, for values of $\omega$ sufficiently small, Eq. (\ref{asymp}) yields the Gaussian density
\begin{equation}
\tilde{p}(\omega,t)\simeq e^{-\mu_2t\omega^2/2\langle\tau\rangle},
\label{gaussian}
\end{equation}
which is equivalent to the central limit theorem result. 

On the other hand for a long-tailed jump density for which $\tilde{h}(\omega)\simeq 1-k|\omega|^{\nu}$ as $\omega\rightarrow 0$ (note that if $0<\nu<2$, $h(x)$ has infinite variance) and assuming that for 
$\omega$ small $\langle\tau e^{i\omega\Delta X}\rangle\simeq\langle\tau\rangle$, we obtain from Eq. (\ref{asymp}) the L\'{e}vy distribution 
\begin{equation}
\tilde{p}(\omega,t)\simeq e^{-k|\omega|^\nu t/\langle\tau\rangle}.
\label{levy}
\end{equation}
The L\'{e}vy distribution has been proposed by several authors as a good candidate for describing financial distributions when the observed fat tails are not Gaussian \cite{mandelbrot,fama,stanley}. However, from the discussion above we see that in the framework of CTRW's the appearance of L\'{e}vy distributions is linked to the existence of transactions for which the jumps have infinite variance. Since this cannot occur in real markets we conclude that the L\'{e}vy distribution is an unsuitable candidate for the propagator.

We finally show that the tails of the return pdf $p(x,t)$ are the same as that of the jump pdf $h(x)$. Indeed, as is well known \cite{weissllibre} the tails of $p(x,t)$ as $|x|\rightarrow\infty$ are determined by the behavior of its characteristic function as $\omega\rightarrow 0$. Thus suppose that $\omega$ is small so that 
$\langle\tau e^{i\omega\Delta X}\rangle\simeq\langle\tau\rangle$ and $t$ is moderate in the sense that the approximation given in Eq. (\ref{asymp}) is still valid but in such a way that $[1-\tilde{h}(\omega)]t/\langle\tau\rangle<1$. Then from Eq. (\ref{asymp}) we see that 
$$
\tilde{p}(\omega,t)\simeq-1[1-\tilde{h}(\omega)]t/\langle\tau\rangle,
\qquad(\omega\rightarrow 0),
$$
which after Fourier inversion and neglecting delta function terms yields
\begin{equation}
p(x,t)\sim\frac{t}{\langle\tau\rangle}h(x),\qquad(|x|\rightarrow\infty).
\label{tailsp}
\end{equation}
Therefore, the analytic form of the tails of $p(x,t)$ as a function of $x$ coincides with that of $h(x)$.

\section{Volatility and inverse problems}
\label{sec4}

Aside from the pdf $p(x,t)$, which provides maximal information about the evolution of $X(t)$, there is another quantity of considerable practical interest: the variance of $X(t)$. In our analysis this quantity, called ``volatility" in the terminology of finance, has the advantage that it does not require the knowledge of the entire jump distribution $h(x)$. It suffices to know the pdf  $\psi(t)$ and the first two moments of $h(x)$.

Let $\langle X^n(t)\rangle$ be the $n$-th moment of the return process:
$$
\langle X^n(t)\rangle=\int_{-\infty}^{\infty}x^n p(x,t)dx,
$$
and let us denote by $\hat{m}_n(s)$ its Laplace transform
$$
\hat{m}_n(s)\equiv\int_0^\infty e^{-st}\langle X^n(t)\rangle dt.
$$
This can be written in terms of the joint Fourier-Laplace transform of $p(x,t)$ by
\begin{equation}
\hat{m}_n(s)=
i^{-n}\left.\frac{\partial^n\tilde{p}(\omega,s)}{\partial\omega^n}\right|_{\omega=0}.
\label{moments1}
\end{equation}
Recall that a direct consequence of the unbiased assumption expressed in Eq. (\ref{even_h}) is that the odd moments of $h(x)$ are equal to zero. Let $\mu_n$ be the $n$th moment of the jump density. Then the $2n$th moment can be written in terms of the $2n$th derivative of the characteristic function as $\mu_{2n}=(-1)^n\tilde{h}^{(2n)}(0)$. All of this implies that the odd moments of the return process vanish:
\begin{equation}
\langle X^{2n-1}(t)\rangle=0,\qquad(n=1,2,3,\cdots).
\label{Xnodd}
\end{equation}

The combination of Eqs. (\ref{hatp}) and (\ref{moments1}) leads to the relation
\begin{equation}
\hat{m}_2(s)=\frac{\hat{R}_2(s)}{s[1-\hat{\psi}(s)]},
\label{volatilitygeneral}
\end{equation}
where 
\begin{equation}
\hat{R}_2(s)\equiv
-\left.\frac{\partial^2\tilde{\rho}(\omega,s)}{\partial\omega^2}\right|_{\omega=0}.
\label{hatR2}
\end{equation}

For the Clayton copula defined by Eq. (\ref{rhocopula}) we see that 
$$
\hat{R}_2(s)=\mu_2\hat{\psi}^{1+\alpha}(s),
$$
where $\mu_2=\langle \Delta X^2\rangle$ is the second moment of the jump density $h(x)$. Hence
\begin{equation}
\hat{m}_2(s)=\mu_2\frac{\hat{\psi}^{1+\alpha}(s)}{s[1-\hat{\psi}(s)]}.
\label{mcopula}
\end{equation}
For Poissonian sojourns this equation yields
$$
\hat{m}_2(s)=\frac{\mu_2\lambda^{1+\alpha}}{s^2(\lambda+s)^\alpha},
$$
which after Laplace inversion reads \cite{erderlyi}
\begin{equation}
\langle X^2(t)\rangle=
\frac{\mu_2}{\Gamma(2+\alpha)}(\lambda t)^{1+\alpha}F(\alpha,\alpha+2,-\lambda t)
\label{volatilitycopula}
\end{equation}
$(\alpha>-2)$, where $F(a,c,x)$ is the Kummer function. Taking into account the following limiting behaviors of the Kummer function \cite{mos}
$$
F(\alpha,\alpha+1,-\lambda t)\sim 1 \quad(t\rightarrow 0)
$$
and
$$
F(\alpha,\alpha+1,-\lambda t)\sim\frac{\Gamma(\alpha+2)}{\Gamma(2)}(\lambda t)^{-\alpha}\quad(t\rightarrow\infty),
$$
we have
\begin{equation}
\langle X^2(t)\rangle\sim\frac{\mu_2}{\Gamma(2+\alpha)}(\lambda t)^{1+\alpha}
\qquad(\lambda t\ll 1)
\label{volcopulaas_0}
\end{equation}
and
\begin{equation}
\langle X^2(t)\rangle\sim\mu_2\lambda t\qquad (\lambda t\gg 1).
\label{volcopulaass}
\end{equation}
We observe an anomalous diffusion behavior at short times and a diffusion behavior at long times. 

In fact one can easily show that the asymptotic volatility grows linearly with time regardless the specific form of $\rho(x,t)$ selected. Indeed, since in all realistic models $\mu_2$ and $\langle\tau\rangle$ are finite, we see from Eq. (\ref{asymp}) that
\begin{equation}
\langle X^2(t)\rangle\sim\frac{\mu_2}{\langle\tau\rangle}\ t\qquad(t\rightarrow\infty).
\label{asympvol}
\end{equation}
Therefore the asymptotic return volatility increases proportional to $t$ independent of the model chosen since 
low-order moments are supposed to be finite in realistic models. Again this is a consequence of the central limit theorem.

\subsection{Inverse problems}

In the above development we have determined the return density $p(x,t)$ after knowing the densities $\psi(t)$ and $h(x)$, we can call this ``the direct problem"'. However, from the CTRW formalism it is possible to infer the mechanism of price formation at a microscopic level by obtaining the statistics of tick-by-tick data represented by the pausing-time density $\psi(t)$ and the return increment density $h(x)$ once we know the volatility $\langle X^2(t)\rangle$ and the return pdf $p(x,t)$. We term this ``the inverse problem". Obviously this problem will require the knowledge of the complete pdf $p(x,t)$ for all times $t$, something that can be beyond our reach, since in practice one may only have access, for instance, to daily data from which one is able to guess the pdf for the daily return $p(x,t)$, where $t=1$ day. In such a case we do not have an analytical expression for $p(x,t)$ which would allow us to get the microscopic densities $\psi(t)$ and $h(x)$ exactly. Nevertheless with the formalism presented below we will be able to guess which forms of $\psi(t)$ and $h(x)$ are consistent with the observed $p(x,t)$. 

We will now outline the solution of the inverse problem and refer the reader to Ref. \cite{jebo} for complementary material on this subject which include examples based in actual cases. 

\begin{figure}
\centerline{\includegraphics[width=12cm]{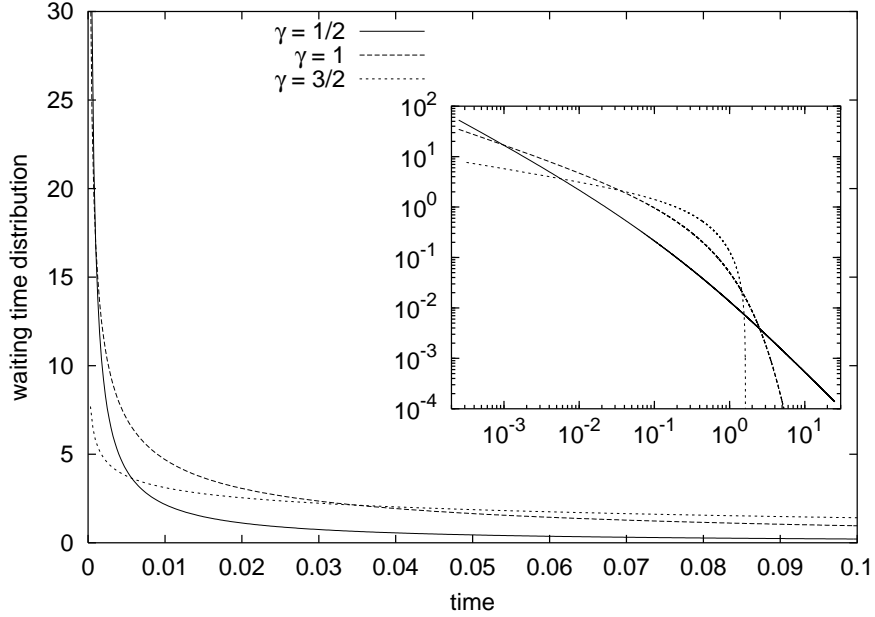}}
\caption{The inverse Laplace transform of the pausing time distribution (\ref{solution_2a}) for $\gamma=1/2, 1, 3/2$ and with time measured in units of $(b/2)^{1/\gamma}t$. We present in the inset the log-log representation of $\psi(t)$.}
\label{psi2}
\end{figure}

Assuming for the joint density $\rho(x,t)$ the functional form given by the Clayton copula, Eq. (\ref{rhocopula}), the formulation to the problem given by Eq. (\ref{formalsolution}) provides a relation between the Fourier and Laplace transforms of the daily return density $p(x,t)$ and tick-by-tick densities $\psi(t)$ and $h(x)$. Obviously a second  relation is required to determine these microscopic densities. This is supplied by the volatility. Indeed, we write Eq. (\ref{mcopula}) as
\begin{equation}
\hat{\psi}^{1+\alpha}(s)+\frac{s\hat{m}_2(s)}{\mu_2}\hat{\psi}(s)-\frac{s\hat{m}_2(s)}{\mu_2}=0,
\label{trascendental_equation}
\end{equation}
which is an algebraic transcendental equation for the now unknown function $\hat{\psi}(s)$. The simplest situation corresponds to $\alpha=-1$, for in such a case Eq. (\ref{trascendental_equation}) gives at once 
\begin{equation}
\hat{\psi}(s)=1-\frac{\mu_2}{s\hat{m}_2(s)}.
\label{inverse_solution}
\end{equation}
Other cases in which a closed solution is readily obtained correspond to $\alpha=-1/2$ and $\alpha=1$. Let us briefly comment the latter case. When $\alpha=1$ Eq. (\ref{trascendental_equation}) reduces to a second degree equation, whose solution reads
\begin{equation}
\hat{\psi}(s)=\frac{s\hat{m}_2(s)}{2\mu_2}\left[\sqrt{1+\frac{4\mu_2}{s\hat{m}_2(s)}}-1\right].
\label{solution_2}
\end{equation}
As an example let us see for which density $\psi(t)$ corresponds to an anomalous diffusion volatility of the form
\begin{equation}
\langle X^2(t)\rangle=kt^\gamma \qquad(k,\gamma>0).
\label{anomalous_diff}
\end{equation}
In this case
$$
\hat{m}_2(s)=\frac{k\Gamma(1+\gamma)}{s^{1+\gamma}},
$$
and from Eq. (\ref{solution_2}) we have
\begin{equation}
\hat{\psi}(s)=\frac{b}{2}s^{-\gamma}\left[\sqrt{1+\frac{4s^\gamma}{b}}-1\right],
\label{solution_2a}
\end{equation} 
where
\begin{equation}
b\equiv k\Gamma(1+\gamma)/\mu_2,
\label{b}
\end{equation}
and $\Gamma(x)$ is the Gamma function. We proceed to the numerical inversion of Eq. (\ref{solution_2a}) by using the Stehfest algorithm \cite{jebo,stehfest}. In Fig. \ref{psi2} we plot the density $\psi(t)$ thus obtained for several values of the exponent $\gamma$. 

We can also obtain, by using Tauberian theorems, the long and short time behaviors of $\psi(t)$. In effect, for small values of $s$, Eq. (\ref{solution_2a}) yields
$$
\hat{\psi}(s)\simeq 1-s^\gamma/b\qquad(s\ll b^{1/\gamma})
$$
and the Tauberian theorems imply that \cite{tauberian}
\begin{equation}
\psi(t)\simeq\frac{\mu_2\sin\pi\gamma}{\pi k}t^{-1-\gamma}\qquad(t\gg b^{-1/\gamma}).
\label{solution_2b}
\end{equation}
On the other hand, for large values of $s$, we have
$$
\hat{\psi}(s)\simeq b^{1/2}s^{-\gamma/2}\qquad(s\gg b^{1/\gamma}).
$$
Hence
\begin{equation}
\psi(t)\simeq\frac{b^{\gamma/2}}{\Gamma(\gamma/2)}t^{-1+\gamma/2}\qquad(t\ll b^{-1/\gamma}).
\label{solution_2c}
\end{equation}
We thus have a power-law pausing density at short and long times. 

Let us finish this section by pointing out that once we have an estimate of $\psi(t)$, we can address the problem of obtaining the pdf of the return increment $h(x)$. Note that setting $s=0$ in Eq. (\ref{hatp}) and taking into account that ({\it cf} Eq. (\ref{hatp0})) 
$\tilde{p}_0(\omega,0)=\langle T\rangle$, we get 
\begin{equation}
\tilde{p}(\omega,0)=\frac{\langle T\rangle}{1-\tilde{h}(\omega)},
\label{hatps0}
\end{equation}
so that 
\begin{equation}
\tilde{h}(\omega)=1-\frac{\langle T\rangle}{\tilde{p}(\omega,0)}.
\label{tildeh}
\end{equation}
Since 
$$
\tilde{p}(\omega,0)=\int_0^\infty\tilde{p}(\omega,t)dt,
$$
Eq. (\ref{tildeh}) allows us to get the characteristic function corresponding to the jump density $h(x)$ if we know the entire return pdf $p(x,t)$ for all $t\geq 0$. In consequence Eq. (\ref{tildeh}) is rather difficult to implement since, as we have mentioned, in many practical situations one only knows at most a numerical series for the value of the propagator instead of an analytic expression of $p(x,t)$. In these cases Eq. (\ref{tildeh}) is useless for evaluating $h(x)$. However, Eq. (\ref{tildeh}) can be very useful for testing any statistical hypothesis, and eventually evaluating unknown parameters, on the form of $h(x)$.

\section{Some generalizations}
\label{sec5}

Up to this point we have developed and applied the CTRW along the lines set up in our previous works \cite{mmw,jebo}. We will now generalize and extend the formalism in two different ways. The first one include the so-called ``overnight effect'' which will allow to extend the formalism to periods of time greater than one day. The second generalization deals with the general equations for the CTRW when non-Markovian effects, due to past correlations, are taken into account. 

\subsection{The overnight effect}

At first sight, the time $t$ appearing in the return density $p(x,t)$ solution of the renewal equation Eq. (\ref{inteq}), it can be any time interval. However, $t=1$ day is, in practice, the upper bound to this time. This is so because the existing formalism does not take into account the overnight effects which cause that the closing price of one day does not usually coincide with the opening price of the following day \footnote{In our previous work \cite{jebo}, in order to have enough statistics from the only time series of prices of a given market, we have artificially forced the closing price of day $n$ to be equal to the opening price of day $n+1$, in other words, we have neglected the overnight effects.}. Some observable  consequences of the overnight effect on return distributions have been adressed in the literature \cite{wood,stoll,bertram}. We, therefore, extend the CTRW by including the overnight effects into the formalism after assuming them to be of a random character. 

Let $S^{(n)}_{\rm open}$ and $S^{(n)}_{\rm close}$ be respectively the opening and closing price of day $n$. Then the zero-mean return at the end of the session of this day is (see Eq. (\ref{x}))
$$
Z_n=\ln[S^{(n)}_{\rm close}/S^{(n)}_{\rm open}]-\langle\ln[S^{(n)}_{\rm close}/S^{(n)}_{\rm open}]\rangle,
$$
and the zero-mean return due to the overnight effect is
$$
Y_n=\ln[S^{(n+1)}_{\rm open}/S^{(n)}_{\rm close}]-\langle\ln[S^{(n+1)}_{\rm open}/S^{(n)}_{\rm close}]\rangle.
$$
Therefore, the detrended return at the beginig of day $N+1$ is given by the sum
\begin{equation}
X_N=\sum_{n=1}^N(Z_n+Y_n).
\label{X_N}
\end{equation}
Suppose now that both $Z_n$ and $Y_n$ are independent and identical distributed random variables with probability density functions given respectively by $p(x)$ and $\varphi(x)$, {\it i.e.},
$$
p(x)dx={\rm Pr}\{x<Z_n<x+dx\},\qquad \varphi(x)dx={\rm Pr}\{x<Y_n<x+dx\};
$$
and let $P_N(x)$ be the probability density function of the return at the beginning of day $N+1$:
$$
P_N(x)dx={\rm Pr}\{x<X_N<x+dx\}.
$$
If $\tilde{P}_N(\omega)$, $\tilde{p}(\omega)$, and $\tilde{\varphi}(\omega)$ denote their corresponding characteristic functions, then Eq. (\ref{X_N}) and the assumption of independent events lead to 
\begin{equation}
\tilde{P}_N(\omega)=\left[\tilde{p}(\omega)\tilde{\varphi}(\omega)\right]^N.
\label{P_N}
\end{equation}
This equation determines the probability distribution of the return at the beginning of day $N+1$ if one knows the distributions $p(x)$ and $\varphi(x)$. In a practical situation the density for an overnight transition, $\varphi(x)$,  has to be guessed from data. Nevertheless, $p(x)$ can be obtained from the CTRW formalism. Indeed, following the notation of the previous sections, let $p(x,t)$ be the return pdf at time $t$ (recall that $p(x,t)$ is the solution of the renewal equation (\ref{inteq})). Then 
$$
p(x)=p(x,t_s)
$$
where $t_s$ corresponds to the length of one session. Since for most markets the average time between successive transaction $\langle\tau\rangle$ is about several seconds and the duration of a session $t_s$ is usually of some hours ($6.5$  hours for the American market), we see that $t_s\gg\langle\tau\rangle$. Therefore we can apply the asymptotic results discussed in Sect. \ref{sec2}, specifically Eq. (\ref{asymp}), with the result
\begin{equation}
\tilde{p}(\omega)\simeq\frac{\langle\tau\rangle}
{\langle\tau e^{i\omega\Delta X}\rangle}
\exp\left\{-\frac{[1-\tilde{h}(\omega)]t_s}{\langle\tau e^{i\omega\Delta X}\rangle}\right\}
\qquad(t_s\gg\langle\tau\rangle),
\label{asymp_N}
\end{equation}
where $\langle\tau e^{i\omega\Delta X}\rangle$ is given by Eq. (\ref{average}). 

Equation (\ref{P_N}) along with Eq. (\ref{asymp_N}) and a guessed form of the overnight jump density $\varphi(x)$ provide a general approximate expression for the probability distribution of returns after $N$ days $(N=1,2,3,\cdots)$. 

Let us finish this section by giving a simple expression for the volatility $\langle X^2_N\rangle$ after $N$ days. In terms of the characteristic function $\tilde{P}_N(\omega)$ this quantity is given by
$$
\langle X^2_N\rangle=\left.-\frac{\partial^2\tilde{P}_N(\omega) }{\partial\omega^2}\right|_{\omega=0}.
$$
From Eq. (\ref{P_N}) and taking into account the assumed symmetry of return jumps, either intraday and overnight, we get
$$
\langle X^2_N\rangle=-N\left.\frac{\partial^2 }{\partial\omega^2}\left[\tilde{p}(\omega,t_s)+
\tilde{\varphi}(\omega)\right]\right|_{\omega=0}.
$$
Let us denote by $\nu_2$ the overnight mean-square return jump:
$$
\nu_2=2\int_0^\infty x^2\varphi(x)dx=-\tilde{\varphi}^{''}(0).
$$
On the other hand, from Eq. (\ref{asymp_N}) we have (see also Eq. (\ref{asympvol})) 
$$
\left.\frac{\partial^2\tilde{p}(\omega,t_s)}{\partial\omega^2}\right|_{\omega=0}\simeq
-\frac{\mu_2 t_s}{\langle\tau\rangle} \qquad(t_s\gg\langle\tau\rangle).
$$
Hence
\begin{equation}
\langle X^2_N\rangle\simeq N\left(\frac{\mu_2 t_s}{\langle\tau\rangle}+\nu_2\right),
\label{N_day_volat}
\end{equation}
where $t_s=6.5$ hours. Therefore, $\langle X^2_N\rangle$ grows linearly with $N$. In other words, the $N$-day volatility ($N=1,2,3,\cdots$) has a normal diffusion behavior \cite{campbell}. 
 
\subsection{A non-Markovian model for the CTRW}

The entire formalism of the CTRW outlined in the preceding sections relies on the fundamental assumption that both waiting times $\tau_n=t_n-t_{n-1}$ and jump increments $\Delta X_n=X(t_n)-X(t_{n-1})$ are independent and identically distributed random variables. However the hypothesis of independence may be questionable. This is specially certain for jump increments in which, in studying certain phenomena related to extreme times \cite{montero_lillo}, there seems to be a degree of dependence between successive jump returns. We will thus generalize our formalism in order to include possible memory effects.

To incorporate these effects the simplest generalization consists in assuming that the joint pdf for return increments and waiting times $\rho(x,t)$ as defined in {\it cf} Eq. (\ref{rhodef}) is replaced by the following Markovian density
\begin{eqnarray}
\rho(x,t|x',t')dxdt&=&{\rm Prob}\{x<\Delta X_n\leq x+dx;\nonumber \\
&\quad & t<\tau_n\leq t+dt|\Delta X_{n-1}=x';\tau_{n-1}=t'\}.
\label{Mrho}
\end{eqnarray}
We remark that in this case the whole process $X(t)$ is always not Markovian because the return pdf $p(x,t|\Delta x_0,\Delta t_0)$ depends on both the magnitude of the previous jump $\Delta x_0=X(t_0)-X(t_{-1})$ and its sojourn time $\Delta t_0=t_0-t_{-1}$; in other words, the return distribution at time $t$ depends on two previous times $t_0$ and $t_{-1}$ ($t<t_0<t_{-1}$). 

Now the integral equation governing the evolution of the return pdf is given by the renewal equation
\begin{eqnarray}
p(x,t|\Delta x_0,\Delta t_0)&=&p_0(x,t|\Delta x_0,\Delta t_0) \nonumber \\
&+& \int_{-\infty}^{\infty}dy\int_0^t d\tau \rho(y,\tau|\Delta x_0,\Delta t_0) p(x-y,t-\tau|y,\tau),
\label{NMrenewal}
\end{eqnarray}
where
\begin{equation}
p_0(x,t|\Delta x_0,\Delta t_0)=\delta(x)\Psi(t|\Delta x_0,\Delta t_0),
\label{NMp0}
\end{equation}
and
\begin{equation}
\Psi(t|\Delta x_0,\Delta t_0)=\int_t^\infty d\tau\int_{-\infty}^\infty \rho(y,\tau|\Delta x_0,\Delta t_0) dy.
\label{NMPsi}
\end{equation}
Note that for the case of independent increments discussed in the previous sections we have 
$$
\rho(y,\tau|\Delta x_0,\Delta t_0)=\rho(y,\tau),
$$
and Eq. (\ref{NMrenewal}) equals Eq. (\ref{inteq}). 

The level of complexity of the above formalism can be somewhat decreased when correlations between consecutive waiting times are negligible. A situation that appears in the study of some financial time series \cite{montero_lillo}. In this case instead of Eq. (\ref{Mrho}) we write
$$
\rho(x,t|x')dxdt={\rm Prob}\{x<\Delta X_n\leq x+dx;t<\tau_n\leq t+dt|\Delta X_{n-1}=x'\},
$$
and the renewal equation for the distribution of $X(t)$ reduces to
\begin{equation}
p(x,t|\Delta x_0)=p_0(x,t|\Delta x_0)+
\int_{-\infty}^{\infty}dy\int_0^t d\tau \rho(y,\tau|\Delta x_0) p(x-y,t-\tau|y).
\label{NMrenewal_2}
\end{equation}
This integral equation can be written in a simpler form using the joint Fourier-Laplace transform:
$$
\tilde{p}(\omega,s|\Delta x_0)=\int_0^\infty e^{-st}dt\int_{-\infty}^\infty e^{i\omega x}p(x,t|\Delta x_0)dx.
$$
We have
\begin{equation}
\tilde{p}(\omega,s|\Delta x_0)=\hat{\Psi}(s|\Delta x_0)+
\int_{-\infty}^{\infty}e^{i\omega y}\hat{\rho}(y,s|\Delta x_0)\tilde{p}(\omega,s|y)dy,
\label{NMrenewal_3}
\end{equation}
where $\hat{\rho}(y,s|\Delta x_0)$ is the Laplace transform of the conditional joint density. We thus see that in this non-Markovian case the Fourier-Laplace transform of the return pdf satisfies an integral equation and, without further assumptions and simplifications, we cannot get a closed expression for it. This contrasts with the Markovian case discussed in Sec. \ref{sec2} in which we got a closed and general expression for $\tilde{p}(\omega,s)$ given by Eq. (\ref{hatp}). 

We finish this section by presenting the integral equation to be satisfied by the volatility. Let 
$$
\hat{m}_n(s|\Delta x_0)=\int_0^\infty e^{-st}\langle X^n(t|\Delta x_0)\rangle dt
$$
be the Laplace transform of the conditional $n$-th moment of the return process:
$$
\langle X^n(t|\Delta x_0)\rangle=\int_{-\infty}^{\infty}x^np(x,t|\Delta x_0)dx.
$$
Hence (see Eq. (\ref{moments1}))
\begin{equation}
\hat{m}_n(s|\Delta x_0)=
i^{-n}\left.\frac{\partial^n\tilde{p}(\omega,s|\Delta x_0)}{\partial\omega^n}\right|_{\omega=0}.
\label{MNmoment}
\end{equation}
From Eqs. (\ref{NMrenewal_3}) and (\ref{MNmoment}) we readily get 
\begin{equation}
\hat{m}_n(s|\Delta x_0)=\sum_{k=0}^{n}\int_{-\infty}^{\infty}y^{n-k}\hat{\rho}(y,s|\Delta x_0)\hat{m}_k(s|y)dy
\label{NMmoment_eq}
\end{equation}
which constitutes a recursive set of integral equations for $\hat{m}_n(s|\Delta x_0)$ $(n=1,2,3,\cdots)$. If we extend to this non-Markovian case the unbiased assumption for the joint density $\rho$:
$$
\rho(x,t|\Delta x_0)=\rho(-x,t|\Delta x_0),
$$
one can easily prove that the odd conditional moments vanish, 
$$
\langle X^{2n-1}(t|\Delta x_0)\rangle=0,
$$
and the Laplace transform of even moments satisfy
\begin{equation}
\hat{m}_{2n}(s|\Delta x_0)=\sum_{k=0}^{n}\int_{-\infty}^{\infty}y^{2(n-k)}\hat{\rho}(y,s|\Delta x_0)\hat{m}_{2k}(s|y)dy,
\label{NMmoment_eq_even}
\end{equation}
($n=1,2,3,\cdots)$. In particular for $n=1$ we see that the Laplace transform of the (conditional) volatility obeys the integral equation
\begin{equation}
\hat{m}_2(s|\Delta x_0)=
\hat{\mu}_2(s|\Delta x_0)+\int_{-\infty}^{\infty}\hat{\rho}(y,s|\Delta x_0)\hat{m}_2(s|y)dy,
\label{MNvolat_eq}
\end{equation}
where
\begin{equation}
\hat{\mu}_2(s|\Delta x_0)\equiv\int_{-\infty}^{\infty}y^2\hat{\rho}(y,s|\Delta x_0)dy
\label{MNmu}
\end{equation}
is the second moment of the return increments.

\section{Conclusions}
\label{sec6}

We have presented a comprehensive review of the CTRW formalism applied to financial systems. This technique is specially suited to deal with  high-frequency data and allows the treatment of both direct problems and inverse problems. We recall that by a direct problem we understand obtaining the ``macroscopic'' pdf, $p(x,t)$, and volatility, $\langle X^2(t)\rangle$, from the ``microscopic'' densities $\psi(t)$ and $h(x)$, while by and inverse problem the situation is reversed, that is, one has to obtain the microscopic densities out of the macroscopic ones. 

We have extended our previous treatment of the joint density of sojourn times and return increments in a more systematic way by using the notion of ``copula''. We have work in detail the so-called Clayton copula. On the other hand, we have introduced into the formalism the overnight effects, which allow us to obtain results for times higher than one day. This will allow to treat more complex inverse problems. This aspect is under current research. 

We have finally generalized the usual formalism, based on the assumption of independent events, to include memory effects. This being motivated by some empirical findings on the existence of longer correlations which indicate the need for a non-Markovian treatment of the problem. We have presented the general setting of a non-Markovian CTRW and in a forthcoming publication we will do a deeper treatment and apply the formalism to actual cases.

\begin{ack}
This work has been supported in part by Direcci\'on General de Investigaci\'on under contract No. FIS2006-05204. 
\end{ack}

\appendix

\section{Derivation of Eq. (22)}
\label{A}

From Eqs (\ref{poissoncase(a)}) and (\ref{poissoncase}), we see that
\begin{equation}
\hat{\psi}(s)=\frac{\lambda}{\lambda+s},\qquad
\tilde{h}(\omega)=\frac{1}{1+\omega^2/\gamma^2}.
\label{poissoncase2}
\end{equation}
Since $\alpha=1$ then the Clayton copula (\ref{rhocopula}) yields
\begin{equation}
\tilde{\rho}(\omega,s)=\frac{\lambda}{s+\lambda(1+\omega^2/\gamma^2)},
\label{rhopoisson}
\end{equation}
or, after inverting, 
\begin{equation}
\rho(x,t)=\gamma\left(\frac{\pi\lambda}{t}\right)^{1/2}e^{-\lambda t-\gamma^2x^2/4\lambda t}.
\label{rhopoisson(b)}
\end{equation}
Equation (\ref{formalsolution}) now reads
\begin{equation}
\tilde{p}(\omega,s)=
\frac{1}{\lambda+s}\left[1+\frac{1}{\omega^2/\gamma^2+s/\lambda}\right]. 
\label{poissoncase3}
\end{equation}
The inverse Fourier transform yields \cite{erderlyi} 
$$
\tilde{p}(x,s)=\frac{1}{\lambda+s}\delta(x)+
\frac{\gamma\sqrt{\lambda}}{2(\lambda+s)\sqrt{s}}e^{-\gamma|x|\sqrt{s/\lambda}}.
$$
Taking into account the following Laplace inversion
$$
{\mathcal L}^{-1}\left\{\frac{1}{\sqrt{s}}e^{-a\sqrt{s}}\right\}=
\frac{e^{-a^2/4t}}{(\pi t)^{1/2}},
$$
and the convolution theorem, we finally obtain the following closed expression for the return pdf 
\begin{equation}
p(x,t)=e^{-\lambda t}\left[\delta(x)+\frac{\gamma}{\sqrt{\pi}}
\int_0^{\sqrt{\lambda t}}e^{\xi^2-\gamma^2x^2/4\xi^2}d\xi\right],
\label{poissonlaplace_app}
\end{equation}
which is Eq. (\ref{poissonlaplace}).

\section{Derivation of Eq. (23)}
\label{B}

Since $\Delta X$ and $\tau$ are random variables representing respectively returns increments and sojourn times, we can write $\tilde{\rho}(\omega,s)$ in the form (see Eq. (\ref{rhodef}))
$$
\tilde{\rho}(\omega,s)=\langle\exp\{i\omega\Delta X-s\tau\}\rangle. 
$$ 
Expanding this average around $s=0$, we get
\begin{equation}
\tilde{\rho}(\omega,s)\simeq\tilde{h}(\omega)-s\langle\tau e^{i\omega\Delta X}\rangle
\qquad(s\rightarrow 0),
\label{asymrho}
\end{equation}
where we have taken into account that the characteristic function for jumps, $\tilde{h}(\omega)$, is given by 
$$
\tilde{h}(\omega)=\langle e^{i\omega\Delta X}\rangle.
$$
On the other hand, if we assume that the sojourn-time density $\psi(t)$ has finite moments  then 
\begin{equation}
\hat{\psi}(s)\simeq 1-s\langle\tau\rangle,\qquad(s\rightarrow 0),
\label{hatpsi0}
\end{equation}
where $\langle\tau\rangle$ is the mean sojourn time. From Eq. (\ref{hatp0}) we have 
\begin{equation}
\tilde{p}_0(\omega,s)\simeq\langle\tau\rangle,\qquad(s\rightarrow 0).
\label{asymp0}
\end{equation}
Substituting Eqs. (\ref{asymrho}) and (\ref{asymp0}) into Eq. (\ref{hatp}) yields
\begin{equation}
\tilde{p}(\omega,s)\simeq\frac{\langle\tau\rangle}
{1-\tilde{h}(\omega)+s\langle\tau e^{i\omega\Delta X}\rangle}\qquad(s\rightarrow 0).
\label{asymhatp}
\end{equation}

By virtue of Tauberian theorems \cite{weissllibre,tauberian}, the asymptotic behavior of the characteristic function $\tilde{p}(\omega,t)$ as $t\rightarrow\infty$ will be given by the inverse Laplace transform of Eq. (\ref{asymhatp}). This inversion reads
\begin{equation}
\tilde{p}(\omega,t)\simeq\frac{\langle\tau\rangle}
{\langle\tau e^{i\omega\Delta X}\rangle}
\exp\left\{-\frac{[1-\tilde{h}(\omega)]t}{\langle\tau e^{i\omega\Delta X}\rangle}\right\}
\qquad(t\rightarrow\infty),
\label{asymp-final}
\end{equation}
which proves Eq. (\ref{asymp}).

\end{document}